\documentstyle[12pt,aaspp4,flushrt]{article}
\makeatletter

\@ifundefined{chapter}{\def\thebibliography#1{\section*{References\@mkboth
  {REFERENCES}{REFERENCES}}\list
  {\relax}{\setlength{\labelsep}{0em}
        \setlength{\itemindent}{-\bibhang}
        \setlength{\itemsep}{0pt}
        \setlength{\parsep}{0pt}
        \setlength{\leftmargin}{\bibhang}}
    \def\newblock{\hskip .11em plus .33em minus .07em}
    \sloppy\clubpenalty4000\widowpenalty4000
    \sfcode`\.=1000\relax}}%
{\def\thebibliography#1{\chapter*{Bibliography\@mkboth
  {BIBLIOGRAPHY}{BIBLIOGRAPHY}}\list
  {\relax}{\setlength{\labelsep}{0em}
        \setlength{\itemindent}{-\bibhang}
        \setlength{\itemsep}{0pt}
        \setlength{\parsep}{0pt}
        \setlength{\leftmargin}{\bibhang}}
    \def\newblock{\hskip .11em plus .33em minus .07em}
    \sloppy\clubpenalty4000\widowpenalty4000
    \sfcode`\.=1000\relax}}

\newlength{\bibhang}
\let\@internalcite\cite
\def\cite{\let\@citeleft(\let\@citeright)%
    \@ifstar{\citeyear}{\citefull}}
\def\citenp{\let\@citeleft\relax\let\@citeright\relax
    \@ifstar{\citeyear}{\citefull}}
\def\citefull{\def\astroncite##1##2{##1~##2}\@internalcite}
\def\citeyear{\def\astroncite##1##2{##2}\@internalcite}

\def\@citex[#1]#2{\if@filesw\immediate\write\@auxout{\string\citation{#2}}\fi
  \def\@citea{}\@cite{\@for\@citeb:=#2\do
    {\@citea\def\@citea{; }\@ifundefined
       {b@\@citeb}{{\bf ?}\@warning
       {Citation `\@citeb' on page \thepage \space undefined}}%
{\csname b@\@citeb\endcsname}}}{#1}}

\def\@cite#1#2{\@citeleft#1\if@tempswa , #2\fi\@citeright}
\def\@biblabel#1{}

\makeatother

\newcommand{\PSbox}[3]{\mbox{\rule{0in}{#3}\includegraphics{#1}\hspace{#2}}}
\newcommand{\FigNum}[1]{\unitlength 1pt \begin{picture}(55,10)(-400,35) 
                        \put(0,0){Figure #1}
                        \end{picture}}
\def\persec{{sec$^{-1}$}}

\def\kmpersec{{km {sec$^{-1}$}}}
\def\approxlt{\lower.2em\hbox{$\buildrel < \over \sim$}}
\def\approxgt{\lower.2em\hbox{$\buildrel > \over \sim$}}

\def\ie{{\it i.e.\/}}
\def\lya{\ifmmode {{\rm Ly}\alpha}
        \else {Ly$\alpha$}\fi}
\def\bbar{\ifmmode {{\not{\rm b}}}
	\else {{${\not{\rm b}}$}}\fi}

\received{July 30, 1997}
\slugcomment{\it Astrophys. J. Letters, submitted}

\begin{document}

\title{Sheets and Filaments as the Origin of the High-Velocity Tail of the \lya\ Forest}
\author{Robert E. Rutledge}
\affil{Max-Planck-Institut f\"ur Extraterrestrische Physik, 
Postfach 1603, D-85740 Garching, Germany 
\\
  rutledge@rosat.mpe-garching.mpg.de}

\begin{abstract}
Simulations of large-scale structure formation predict the formation
of sheet- and filamentary structures, which are often invoked as the
origin of the \lya\ forest.  In their simplest description, these
sheets and filaments require a differential distribution of observed
line-of-sight velocity widths ($b$) which will decrease as power-laws
at velocities well above the observed peak in this distribution: for
filaments, the differential distribution is $dN/db \propto b^{-3}$,
while for sheets it is $dN/db \propto b^{-2}$.  These functional
dependences on $b$ arise {\it a priori} due to the geometry of these
absorbing structures -- assuming random orientations relative to the
line-of-sight -- and are otherwise unrelated to the physical state in
the absorbing structure.  We find the the distribution at $b>35$ in
three previously published data sets to be steeper than $dN/dB \propto
b^{-3}$ (99.99\% confidence).  This implies that evidence of the finite
length of these kinds of absorbing structures is present in the
$b$-distribution data.

\end{abstract}


\section{Introduction}

The observational distribution of measured velocity widths ($b$) of
\lya\ forest absorption lines may be parameterized as Gaussian, but
with a significant non-Gaussian, high-velocity tail ($\ge 55$
\kmpersec; cf. \citenp{hu95,lu96,kim97,kirkman97,khare97} for recent observational results). 

Investigations based on spherical gaseous clouds bounded by external
pressure \cite{duncan91,petitjean93} find that the $b$-distribution
can only be reproduced by a population with a strong correlation
between central density, radii, and cloud mass, requiring a tight
correlation between column density and $b$, which has not been
strongly supported by the data (\citenp{rauch93}, discusses selection
effects which may have produced earlier reported correlations).  Other
researchers find that high implied temperatures of such clouds
(assuming $b=\sqrt{2kT/m_p}$) cannot be produced by structures in
thermal equilibrium with an ionizing UV background, and may imply
recent adiabatic collapse
\cite{press93}.

The ramifications of CDM structure formation models as the origin of
the \lya\ forest have been extensively investigated.  Recent workers,
using hydrodynamic simulations of cosmological models (usually CDM or
$\Lambda$CDM), describe the structures responsible for the low-density
($N_H \approxlt 10^{15}$) \lya\ forest to be filaments, pancakes,
``cigar-like'', mini-pancakes, or sheets
\cite{cen94,zhang95,hernquist96,katz96,jordi96,bi97,cen97,nath97,zhang97}.
Commonly, the $b$-distribution is parameterized as a Gaussian; in
addition, a non-Gaussian high-velocity tail is universally noted.  The
measured $b$-values do not always correspond directly to a
temperature, and most structure formation theories produce
fluctuations in the density of the inter-galactic medium (IGM) larger
(in redshift space) than the thermal width (\citenp{jordi96,hui96b}).

In this {\it Letter}, we describe how this high-velocity tail may be
generated by the same structures responsible for the lower-velocity
Gaussian peak. The low-velocity (Gaussian peak) values are caused by
the line-of-sight passing nearly perpendicular to the long-axis of the
sheet/filament-type structures, while the high-velocity tail is due
to higher angles of incidence. The relative number of low vs. high
$b$-values are, in this scenario, determined by the geometry of the
absorbing structures, and is otherwise unrelated to the physical state
of the absorbing structure.
 
In exploring this scenario, we keep in mind the fact that the
fluctuations which are produced in large-scale structure formation
simulations are in a continuous density field, and ascribing a
simplified geometry to them in the way we do (as a cylinder, or an
infinite sheet) is an idealization of a structure which has already
been observed in the simulations to be more complicated.

\section{The Model: Line-of-Sight $b$-Distribution}

In this model, the magnitude of $b$ corresponds to the line-of-sight
width of the absorbing structure in velocity space, which is
determined both by its physical size in comoving space and by its
peculiar velocity structure.

\subsection{Cylindrical Filaments}

In the simplest manifestation (refer to Fig.~\ref{fig:absorption}), an
observer's line-of-sight (LOS) crosses an absorber which has a minimum
(velocity) width $b_0$, and which has a long-axis.  We assume that the
geometry of the absorber is such that the cross section -- which is
the plane shared by the short-axis $b_0$ and the LOS, parallel to the
long axis -- can be described as a rectangle of width $b_0$ and
quasi-infinite length.  This is a fair assumption to describe
cross-sections of filaments and sheets, but not of spheres.
Presently, we treat the long-axis as infinite, but the expected finite
length would produce observationally important effects, which we
discuss below.

The LOS crosses the absorber at an angle $\theta$ to the
perpendicular of $b_0$, producing a line-of-sight crossing distance
(in velocity space) of $b$.  The observed velocity-width $b$ is
geometrically related to the short axis $b_0$:

\begin{equation}
\label{eq:b}
b = \frac{b_0}{\sin(\theta)}
\end{equation}

This is the relationship between the LOS velocity-width ($b$) of the
observed absorption line and the velocity-width $b_0$ of the density
enhancement (that is, filament or sheet).  If we assume a random
distribution in $\theta$ (that is, a random orientation of absorbing
structures relative to the LOS), because $\theta=0$ is the axis of
symmetry for the cylindrical filament, the number of LOS per unit
$\theta$ is:
\begin{equation}
\label{eq:cyltheta}
\frac{dN_{\rm cyl}}{d\theta} = \sin(\theta)
\end{equation}

\noindent and the number of LOS per unit velocity for a constant $b_0$ -- is:
 
\begin{equation}
\label{eq:dndb}
\frac{dN_{\rm cyl}}{db} (b, b_0 ) = \frac{dN}{d\theta} \frac{d\theta}{db} = \left\{ \begin{array}{ll}
	- \frac{\left(\frac{b_0}{b}\right)^3}{b_0
\sqrt{1-\left(\frac{b_0}{b}\right)^2}} & b>b_0 \\
	0 & b\le b_0
	\end{array} \right.
\end{equation}

We assume a cylindrical filament, with a diameter $L$; $b_0$ then
depends on an angle $\phi$, corresponding to an impact parameter
relative to the center of the cylinder's circular cross-section:

\begin{equation}
b_0 = L \cos(\phi)
\end{equation}

\noindent where $\phi=0$ when the LOS passes through the center of the
cylinder's circular cross-section, and $\phi\approx \pi/2$ when the
LOS passes nearly tangent to the cylinder's circular cross-section.
The resulting distribution of $b_0$, assuming a random distribution of impact
parameters,  is:

\begin{equation}
\frac{dN}{db_0} = \frac{-\left(\frac{b_0}{L}\right)}{L \sqrt{1 -
\left(\frac{b_0}{L}\right)^2}}
\end{equation}

If we assume a random distribution of angles, the total
$b$-distribution, using Eqn.~\ref{eq:dndb}, becomes

\begin{eqnarray}
\frac{dN_{\rm cyl}}{db} (b,L) = \frac{dN_{\rm cyl}}{db}(b, b_0) \frac{dN}{db_0}\; db_0 &=& \frac{1}{L} \left(\frac{L}{b}\right)^3
	\int^{\frac{\pi}{2}}_{\phi'} \frac{\cos^3 (\phi)}{\sqrt{1 -
	\left(\frac{L}{b}\right)^2 \cos^2(\phi)}} \;  d\phi \\
	\phi' &=& \left\{\begin{array}{ll}
	 0  & b > L   \\
	 \arccos\left(\frac{b}{L}\right) &  b < L
	        \end{array} \right. 
\end{eqnarray}

After making a change of variables of $X=\sin(\phi)$, converting the
integrating variable from $d\phi$ to $d(\sin(\phi))$, and replacing
$\frac{b}{L}= \bbar$, the integral is in a tabulated form
(\citenp{dwight}, 200.01 and 202.01).  Once integrated, we find the
distribution of velocity-widths $b$ to be:
\begin{equation}
\label{eq:cyldndb}
\frac{dN_{\rm cyl}}{db}(b,L) = \left\{ \begin{array}{ll}
	 \frac{1}{2L}\left[ (1 + \frac{1}{\bbar^2})
\ln\left(\frac{\bbar + 1}{\sqrt{\bbar^2 - 1}}\right) -
\frac{1}{ \bbar }\right]  & {\bbar} >  1 \\
     	\frac{1}{2L}\left[ (1 + \frac{1}{\bbar^2})
\ln\left(\frac{\bbar + 1}{\sqrt{1 - \bbar^2}}\right) -
\frac{1}{\bbar}\right]  &  {\bbar} <  1 
	\end{array} \right. 
\end{equation}

This is the normalized $b$-distribution function for an ensemble of
cylindrical filaments of velocity-diameter $L$, oriented randomly with
respect to the line-of-sight.  This distribution function is plotted
in Fig.~\ref{fig:bdist}a, super-imposed on combined datasets 1+3+5 (see
Table~\ref{tab:datasets}).  The distribution is sharply peaked, with a
broader wing at lower-velocities than at higher velocities.  It is a
very poor description of the data near the Gaussian peak, indicating
that, at the very least, a distribution of $L$ values must be imposed.

Expanding the logarithm in \bbar\ in Eq.~\ref{eq:cyldndb}, we find: 
\begin{equation}
\frac{dN_{\rm cyl}}{db}(b,L) \propto \frac{1}{\bbar^3} + \frac{1}{O(b^5)} +
\cdots \;\; \bbar>1 
\end{equation}

\noindent plus higher order terms.  Thus, the differential distribution
of LOS velocities at values greater than the velocity diameter of the
cylinder decreases as an inverse cube of the measured velocity.
Because this decrease is a power-law, integrating over a distribution
in $L$ will not change it's power-law nature at $b > L_{\rm max}$
(where $L_{\rm max}$ is the largest contributing $L$ value).

The differential distribution can be integrated, producing the 
 the normalized cumulative distribution:
\begin{equation}
N_{\rm cyl}(>b;b,L) = \left\{ \begin{array}{ll}
	\frac{1}{2}(1.0 - (\bbar-\frac{1}{\bbar}) \coth{^-1}(\bbar)) & {\bbar} > 1  \\	
	\frac{1}{2}(1.0 - (\bbar-\frac{1}{\bbar}) \tanh^{-1}(\bbar)) &	{\bbar} < 1 
	\end{array} \right.
\end{equation}

\noindent which is the fraction of $b$ values expected to be larger
than some value $b$, for a known value of $L$.  The median is
$\bbar=1.0$ (that is, $b=L$).  Approximately eighty per cent of the
$b$ values are between 0.5 and 2.0 of the median.  At the high
velocity end ($\bbar \gg 1$), the cumulative $b$-distribution
decreases as $\bbar^{-2}$, as expected.  We show this cumulative
distribution for an assumed $L=28$ \kmpersec, super-imposed on
datasets 1+3+5 (see Table~\ref{tab:datasets}), in
Fig.~\ref{fig:bdist}b.  The power-nature of the cumulative
distribution at $b>$28 \kmpersec\ is apparent, and is comparable to
that seen in the dataset.  (We comment on the possibility of a break
in the data near $b=80$ \kmpersec\ in Sec~\ref{sec:break}, where we
statistically compare the observed distribution with a power-law
distribution).

\subsection{Sheets}

The differential distribution for sheets may be found similarly to
cylinders.  The relationship between $b$ and $b_0$ (the short axis
velocity) is as in Eq.~\ref{eq:b}, where again we refer to
Fig.~\ref{fig:absorption}.  However, the axis of symmetry for the
sheet is now perpendicular to the surface of the sheet, and the number
of LOS per angle $\theta$ changes accordingly (compare with Eq.~\ref{eq:cyltheta}):

\begin{equation}
\frac{dN_{\rm sheet}}{d\theta} = \cos(\theta)
\end{equation}

\noindent This is the important difference with the filament
distribution.  We assume a delta-function distribution of $b_0$(=$L$
-- the width of the sheet).  By chain rule, the differential
$b$-distribution is:

\begin{equation}
\frac{dN_{\rm sheet}}{db}(b,L) = \left\{\begin{array}{ll}
			\frac{1}{L} \left(  \frac{L}{b} \right)^2
			& b\ge L \\
			0 &  b< L
	\end{array} \right.
\end{equation}

\noindent resulting in a cumulative distribution: 

\begin{equation}
N_{\rm sheet}(>b; b, L) = \left\{\begin{array}{ll} 
				\frac{L}{b} & b\ge L \\
				1    		  & b< L
			\end{array} \right.
\end{equation}

Thus, the differential distribution expected from sheets is $\propto
b^{-2}$, and the cumulative distribution is $\propto b^{-1}$.  The
differential distribution is non-zero only at $b>$L -- and thus
is truncated at the sheet-width. 

\section{Comparison with Observational Data}

The dependence of the $b$ distribution $dN/db\propto b^{-3}$ or
$b^{-2}$ may be present in existing data-sets.  In this section, we
perform a few comparisons between the data and power-law $b$-distributions.

These comparisons may be hampered by contributions to
line-measurements of line-blending (when 2 or more individual lines
cannot be statistically decomposed, and thus the properties of only
one, broader, blended line is measured), and also to the sensitivity
of the detection measurement algorithms to lines of these velocity
widths.  We have assumed that the measurement algorithms are 100\%
efficient.

We used data from three recently published studies of the \lya\ forest
absorbers \cite{hu95,lu96,kirkman97}.  From these, we have used only
lines with measured column densities $N_H = 10^{12.5}-10^{14}$, and
$b>35$ \kmpersec.  Using a KS-test \cite{press}, we determine the
probability that data from these observations are consistent with
having been drawn from a differential distribution with a
single-power-law, for a range of power-law values in 0.05 increments.
We tested each data-set individually (datasets 2, 4, and 6, see
Table~\ref{tab:datasets}), producing 99.5\% confidence limits (where
the dataset has a 0.5\% probability of being drawn from a parent
distribution with that power-law value).  The results of these are
shown in Table~\ref{tab:datasets}.  The combination of all three
datasets (dataset 7) places limits on a single-value-power-law parent
population for the differential distribution with a power-law between
[$-4.05,-3.15$].  The combined data-set is inconsistent with a
single-value power-law $dN/db \propto b^{-3}$ at 99.99\% confidence,
and inconsistent with $dN/db \propto b^{-2}$ at ($100-2\times 10^{-29}$)\%
confidence.

In Fig.~\ref{fig:bdist}b, there is what looks like an apparent break
from a single power-law in the data near $b=80$ \kmpersec .  In a
combined dataset from all three observations, using values of $b>80$
\kmpersec, we find the 99.5\% upper-limit on the power-law slope is
$-3.5$ (the lower-limit is $-11$).  This is consistent with the range
found for $35<b<60$ \kmpersec ([-4.55,-2.05], 99.5\% confidence range,
drawn from dataset 7; 346 lines).  Thus, the break is not
statistically significant.
\label{sec:break}

\section{Discussion and Conclusions}

We have described a model in which we have assumed the absorbing
structures of the \lya\ forest may be idealized as infinite sheets and
cylinders in velocity space.  In this model, the measured LOS velocity
width $b$ is due to the line-of-sight size in velocity space of the
absorbing structure, determined both by its physical size in comoving
space and by its peculiar velocity structure.  Assuming a random
orientation of these structures relative to the LOS, a
high-velocity-tail to the distribution of LOS velocity widths of the
\lya\ forest is produced.  This tail is due to the chance occurrence
of a LOS passing close to the long-axis, making the total traverse
distance much longer than that of a LOS which passes nearly
perpendicular to the long axis.  The differential distribution due to
an ensemble of infinite cylinders (sheets) shows a dependence $N_{\rm
cyl}(>b) \propto b^{-3}$ ($N_{\rm sheet} (>b) \propto b^{-2}$).

The shape of the $L$-distribution (the characteristic size of the
structure) will dominate the observational $b$-distribution at low
values (\ie\ $b\approx L$).  Thus, in this model the Gaussian which is
observed in the velocity range $b\approx$10-40 km\persec\ is largely
due to the shape of the $L$ distribution, while the high-velocity tail
is due to the line-of-sight penetrating absorbing structures close to
their long-axis.

The single power-law differential $b$-distribution of $b^{-3}$ for
cylinders and $b^{-2}$ for sheets should continue up to values of
$b\sim \infty$ only if these structures are infinitely long.  This is
almost certainly not the case.  At high values of $b$, the
distribution should break from these power-laws, due to the finite
size of the absorbing structures.

A comparison with a combined dataset of 447 lines with $b>$35
\kmpersec\ places limits on the value of the power-law distribution,
with $dN/dB \propto b^{-[4.05,3.15]}$ (99.5\% confidence limits).  A
pure power-law $\propto b^{-3.0}$ is excluded at a probability of
$1\times 10^{-4}$ while $\propto b^{-2.0}$ is excluded at
$2\times10^{-31}$.  Under this model, including values of $b$ toward
lower velocities ($b\sim L_{\rm max}$) could bias the power-law toward
more negative values; however, using a dataset consisting only of
higher velocities ($b>80$ \kmpersec ) the data are even more
inconsistent with $dN/db\propto b^{-3}$. Therefore such low-$b$
contamination cannot be responsible for the inconsistency between the
observational data and the predicted power-law dependency. The effects
of finite-length sheets and filaments may be responsible for the
steepness of this power-law.  Further work on the observational limits
of such a break could provide a direct measurement of the size of
these structures.  Hydrodynamic simulations can play an important role
in predicting the form and magnitude of this break, providing
observers with the tools they need to find it.

In conclusion, modeling the absorbing structures of the IGM
responsible for the \lya\ forest as infinite cylindrical filaments and
sheets in velocity space predicts power-law $b$-distributions -- which
predict more observed high-$b$ lines than are reported in the
observations.  This can be due to several causes, among them the
finite-size of such structures, a decreasing detection efficiency with
increasing $b$, or that the geometric idealization here presented
does not accurately reflect the physical system.

\acknowledgements

RR acknowledges support through a Max-Planck Fellowship from
Max-Planck-Institut f\"ur extraterrestrische physik.  We are extremely
grateful to Lam Hui, for very useful discussions and comments on
drafts. This research has made use of the NASA/IPAC Extragalactic
Database (NED) which is operated by the Jet Propulsion Laboratory,
California Institute of Technology, under contract with the National
Aeronautics and Space Administration.

\newpage

\newpage

\begin{figure} 
\caption{ \label{fig:absorption}       
Schematic diagram of the absorption.  The line-of-sight crosses an
absorber with a minimum size $b_0$ and an infinite length, at an angle $\theta$ with respect to
the perpendicular, producing a total line-of-sight crossing distance
of $b$ \kmpersec.  } 
\end{figure}

\begin{figure} 
\caption{ \label{fig:bdist}       
Comparison of combined datasets 1+3+5 (
Table~\protect{\ref{tab:datasets}}) with theoretical $b$-distribution
due to a cylinder with a single radius $L$=28 \kmpersec . {\bf (a)}
The differential distribution -- the solid histogram is the data, and
the broken line is the theoretical prediction, which is a very poor
approximation to the data; this indicates that a distribution of
cylinder radii is required.  {\bf (b)} Integral of the data (and
theoretical model) shown in panel (a). It is shown in log-log, to make
the power-law nature of both the theoretical distribution and the
observed data clear.  The apparent break near $b=80$ \kmpersec\ in the
data is not statistically significant (see Sec.~\ref{sec:break}). }
\end{figure}

\newpage
\begin{deluxetable}{lcccc}
\tablewidth{33pc}
\tablecaption{$b$ Data-sets \label{tab:datasets}
 }
\tablehead{
\colhead{Dataset} 	& \colhead{$N_H$ range}	& 
\colhead{$b$ range}	& \colhead{Number of}  	& 
\colhead{$dN/db \propto b^{\alpha}$}	\nl
\colhead{\# (ref)}	& \colhead{(cm$^{-2}$)}	& 
\colhead{(\kmpersec)} 	& \colhead{$b$ values}	& 
\colhead{(99.5\% limits)}}
\startdata
1(A)	& $10^{12.5-14}$& --		& 328 		& --		\nl
2(A)	& $10^{12.5-14}$& $>$35		& 116   	& [$-$4.1,$-$2.50]\nl
3(B)	& $10^{12.5-14}$& --    	& 790          	& --		\nl
4(B)	& $10^{12.5-14}$& $>$35 	& 252          	& [$-$4.55,$-$3.05]	\nl
5(C)	& $10^{12.5-14}$& --    	& 287     	& --		\nl
6(C)	& $10^{12.5-14}$& $>$35 	& 84   		& [$-$4.70,$-$2.40]	\nl
7(A+B+C)& $10^{12.5-14}$& $>$35 	& 447  		& [$-$4.05,$-$3.15]	\nl
8(A+B+C)& $10^{12.5-14}$& $>$80 	& 46   		& $<$ $-3.7$	\nl
\tablerefs{
(A) \citenp{kirkman97}: QSO HS 1946+7658, (z=3.02) \nl
(B) \citenp{hu95}:  Q0014+813 (z=3.38), Q0302-003 (z=3.29)
0636+680 (z=3.17), 0956+122 (z=3.30);  
(C) \citenp{lu96}: Q0000-26 (z=4.1) \nl
}
\enddata
\end{deluxetable}

\clearpage
\pagestyle{empty}
\begin{figure}
\PSbox{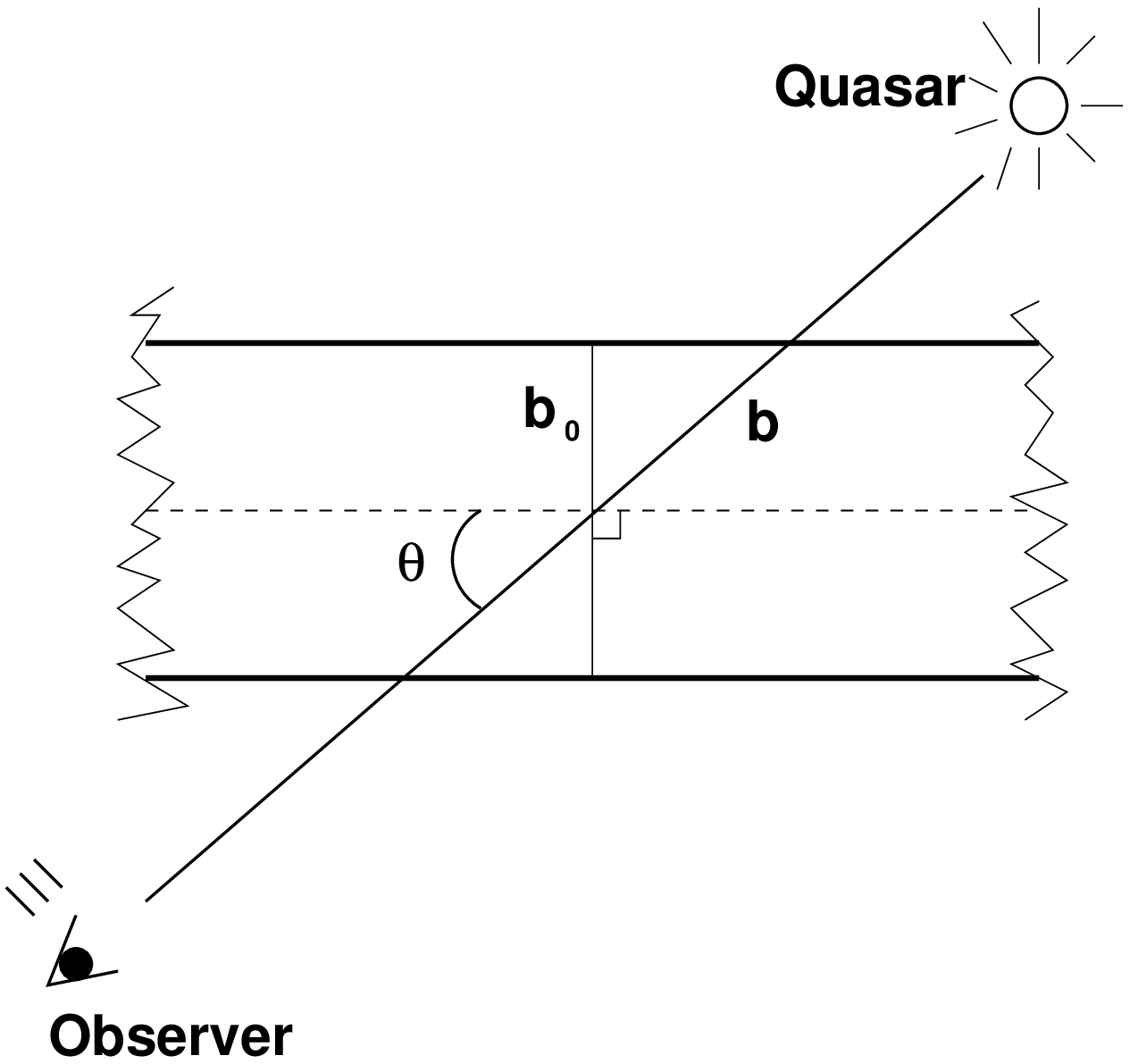 hoffset=0 voffset=0}{14.7cm}{21.5cm}
\FigNum{\ref{fig:absorption}}
\end{figure}

\clearpage
\pagestyle{empty}
\begin{figure}
\PSbox{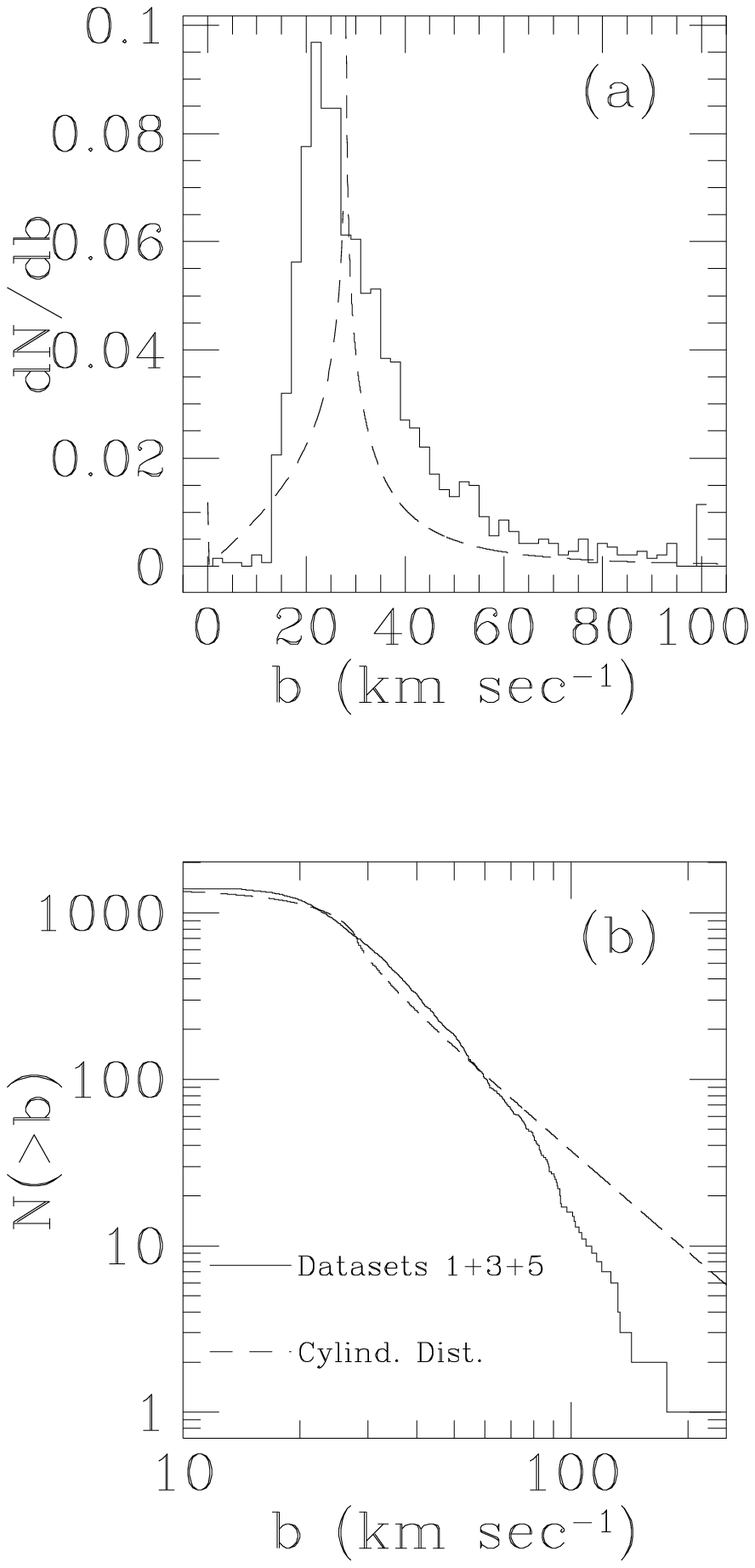 hoffset=0 voffset=-56}{14.7cm}{21.5cm}
\FigNum{\ref{fig:bdist}}
\end{figure}

\newpage

\end{document}